\documentclass{article}
\usepackage{amssymb}
\usepackage{amsfonts}
\usepackage{graphics}
\begin{document}

\title{The Time Periodic Solution of the Burgers Equation \\
on the Half-Line and an Application to Steady Streaming}         
\author{A.S. Fokas$^1$ and J.T. Stuart$^2$ \\
$^1$Department of Applied Mathematics and Theoretical Physics \\
University of Cambridge, \\
Cambridge, CB30WA, UK \\
t.fokas@damtp.cam.ac.uk \\
$^2$Department of Mathematics \\
Imperial College\\
London, SW72BZ, UK\\
t.stuart@imperial.ac.uk}
\date{October 2004}          
\maketitle

\pagestyle{plain}

\begin{abstract}
The phenomenon of steady streaming, or acoustic streaming, is an
important physical phenomenon studied extensively in the
literature. Its mathematical formulation involves the
Navier-Stokes equations, thus due to the complexity of these
equations is usually studied heuristically using formal
perturbation expansions. It turns out that the Burgers equation
formulated on the half line provides a simple model of the above
phenomenon. The physical situation corresponds to the solution of
the Dirichlet problem on the half-line, which decays as $x
\rightarrow \infty$ and which is {\it time periodic}. We show that
the Dirichlet problem, where the usual prescription of the initial
condition is now replaced by the requirement of the time
periodicity, yields a well posed problem. Furthermore, we show
that the solution of this problem tends to the ``inner'' and
``outer'' solutions obtained by the perturbation expansions.
\end{abstract}

\section{Introduction}       
The phenomenon of "steady streaming", or "acoustic streaming" as
it is sometimes called, dates back to the 19th century with the
work of Kundt on the circulations of air in tubes (see Rayleigh
1883).  There are related phenomena of circulations in the flows
under water waves with a free surface (Longuet-Higgins 1953, 1960;
Hunt and Johns 1963).  Following Rayleigh, Schlichting (1932) did
both experimental and theoretical work on a particular case of the
flow induced by a circular cylinder oscillating along a diameter.

The nature of the phenomenon, which is common to all the papers
mentioned above, may be explained as follows. If the flow of fluid
(liquid or gas) is periodic in time but with zero mean, and is
parallel to a rigid surface, then the governing equations ensure
that a thin periodic boundary layer is formed in the neighborhood
of that surface.  If the time frequency is $\omega$-radians per
second and $\nu$ is the kinematic viscosity then the thickness of
this layer is of order $(\nu/w)^{1/2}$.  This boundary layer is
known as a Stokes layer.  A case of particular simplicity is that
in which the flow is parallel to a flat rigid surface, whose
velocity is $U_w\cos w t$ parallel to itself, $t$ being the time
and $U_w$ a constant.  Then the velocity in the fluid neighboring
the surface is
$$u = U_we^{-\eta} \cos(\omega t -\eta), \eqno (1.1)$$
where
$$ \eta = y(w/2\nu)^{1/2}, \eqno (1.2)$$
$y$ being the coordinate normal to the surface.  This is the
prototype Stokes layer (Stokes 1851), see also (Stuart 1963,
Benney 1964). If a Galilean tranfsormation is imposed such that
the surface is at rest but the fluid has velocity $U \cos wt$
parallel to that surface as $y$ and $\eta$ tend to infinity, then
the velocity in the fluid is
$$ U[\cos wt - e^{-\eta} \cos (wt-\eta)]. \eqno (1.3)$$

In  the above description, the fluid motion is intrinsically
linear so that the simple form of the Stokes layer emerges
exactly: the nonlinear terms are identically zero for (1.1) and
(1.3).

The situation is quite different, however, if the velocity (as $y$ and $\eta$ tend to infinity) has the form
$$ U(\xi)\cos wt, \eqno (1.4)$$
where $\xi$ is the coordinate parallel to the surface: the latter
may be curved and its curvature is neglected if the Stokes layer
thickness $(\nu/w)^{1/2}$ is small compared with a typical length
d: thus $\nu/wd^2$ is supposed to be small.  The nonlinear term in
the Navier-Stokes equation, namely $u \cdot \nabla u$, where $u$
is the vector velocity, is not zero however.  It has terms of two
types: (i) there are terms proportional to $\exp(2iwt)$ and to
$\exp(-2iwt)$ which generate corresponding flow components; the
equations that govern these components do have solutions which
tend to zero at the edge of the Stokes layer as is required, (ii)
a mean term is also generated, which drives the steady streaming;
we can refer to this mean term, which is an effect of
rectification, as a Reynolds stress (or, rather a derivative of
the Reynolds stress).  An equation can be obtained for the mean
flow generated by this Reynolds stress but, somewhat
paradoxically, it is not possible to obtain a solution whose
component of velocity that is parallel to the surface, tends to
zero as the edge of the Stokes layer is approached. Rather the
best that can be achieved is to ensure that the solution is finite
for that component of velocity.

Indeed it is found that, as the edge of the Stokes layer is approached, the velocity component approaches
$$ \frac{-3}{4w} U(\xi) \frac{dU(\xi)}{d\xi}. \eqno (1.5)$$

Since $\xi$ has a typical length $d$, this velocity component has
scale $U^2_0/wd$, where $U_0$ is the scale of $U(\xi)$. Now in
problems of the Navier-Stokes equations, the famous Reynolds
number plays a significant role; it is the product of a velocity
and a distance, divided by the kinematic viscosity.  In the
present case we have
$$ R_s = (U^2_0/wd)(d/\nu) = U^2_0/w\nu. \eqno (1.6)$$
This parameter is known as the steady-streaming Reynolds number.  Stuart (1963, 1966) showed the importance of this concept for the calculation of the flow forced by (1.5) outside the Stokes layer.

In the papers of Rayleigh (1883) and Schlichting (1932), it is
implicit that the parameter $R_s$ is small, and the calculation of
the flow outside the Stokes layer is performed on that basis. Even
so, Schlichting's paper makes it clear that he was aware that his
theory for small values of $R_s$ is inappropriate as an
explanation of his experimental work, for which $R_s =250$.

Stuart (1963, 1966) showed how an "outer-boundary-layer theory", which is valid for large values of $R_s$, can be used to obtain a solution for which (1.5) applies at the edge of the Stokes layer but which tends to zero as the distance form the surface tends to infinity.  Implicit in that work is the idea that this "outer" (or "steady-streaming') boundary layer is much thicker than the Stokes layer, indeed by a factor $R_s^{1/2}$.  Following Stuart's work, Riley (1998) has pursued problems of this type in much detail.

A problem can be posed that shows many of the characteristics that
are outlined above, but without the complications of the full
Navier-Stokes equations.  The relevant equation is a Burgers type
equation for $u(x,t)$, namely
$$u_t + \beta (u-k \beta)u_x = \frac{1}{2} u_{xx}, \eqno (1.7)$$
where $\beta$ is a small parameter which plays a role analogous to $R_s^{-1/4}$ above, $R_s$ being large and $k$ is $O(1)$.  The boundary conditions are
$$x =0: \quad u(0,t)=\cos t,$$
$$ x \rightarrow \infty: \quad u(\infty,t) \rightarrow 0, \eqno (1.8)$$
and
$$ u(x,t) = u(x,t+2\pi).$$
The Cole-Hopf transformation
$$u-k\beta = -\theta_x/\beta\theta \eqno (1.9)$$
gives
$$ \theta_t - \frac{1}{2}\theta_{xx} = 0, \eqno (1.10)$$
with the boundary conditions
$$ x=0: \quad \theta_x(0,t) + \beta(\cos t- k\beta)\theta(0,t)=0,$$
$$ x\rightarrow \infty: \quad \theta_x(\infty,t) / \theta(\infty, t) = k\beta^2, \eqno (1.11)$$
$$ \theta_x(x,t)/\theta(x,t) = \theta_x(x,t+2\pi)/\theta(x,t+2\pi).$$
The equation (1.10) is simply stated, but the boundary and periodicity conditions are complicated.  We return to a discussion of this problem in later sections.

In the meantime we note that a formal solution to (1.7), (1.8) can be obtained by expanding $u$ in a power series in $\beta$
$$u=u_0(x,t) +\beta u_1(x,t) + \beta^2u_2(x,t)+\cdots,$$
and it is quickly found that
$$ u_0 = e^{-x}\cos(t-x). \eqno (1.12)$$
At order $\beta$, the $u_1$ terms involving $e^{2it}$ and
$e^{-2it}$ give no difficulty, but the steady (or rectified) term
has to satisfy
$$u_{1xx} =-e^{-2x};$$
the solution which is bounded as $x\rightarrow \infty$ and which is zero at $x=0$ is
$$u_1 = \frac{1}{4} (1-e^{-2x}). \eqno (1.13)$$
We note that as $x\rightarrow \infty$, $u\rightarrow
\frac{1}{4}\beta$ plus higher order terms.  A re-scaling of (1.7)
for the steady part of the solution with
$$ u(x) = \beta v(x), \quad x = z/2\beta^2$$
yields
$$ (v-k)v_z = v_{zz}, \eqno (1.14)$$
with the boundary conditions
$$ z\rightarrow 0: \quad v\rightarrow 1/4, $$
$$ z\rightarrow \infty: \quad v\rightarrow 0. \eqno (1.15)$$
The solution of (1.14) subject to (1.15) is
$$ v = 2k/[(8k-1)e^z +1]. \eqno (1.16)$$

We note that the length scale of this "outer" region is
$(2\beta^2)^{-1}$ times the scale of the "inner" region of (1.12)
and (1.13). Also $z\rightarrow 0$ in (1.15) corresponds to $x
\rightarrow \infty$ in (1.13), in the sense of $\beta \rightarrow
0$ with $x$ fixed and $\beta \rightarrow 0$ with $z$ fixed giving
an equivalence.

This heuristic argument will be justified in later sections by
solving explicitly Burgers equation (1.7) with the conditions
(1.8). We note that the mathematical novelty of this problem is
that it is posed on the half-line and that it requires periodicity
in $t$. We recall that the usual Dirichlet problem for the Burgers
equation on the half-line was analysed by Calogero and De Lillo.
In their formulation, in addition to the Dirichlet boundary
condition, one also specifies the initial condition $u(x,0)$. In
the present situation, instead of specifying $u(x,0)$, one
requires periodicity.

In section 3 we will analyse the time periodic solution of the
Burgers equation on the half-line with two different boundary
conditions at the origin: we specify either (a) the integral of
$u(x,t)$, or (b) $u(0,t)$. The physical problem corresponds to the
case (b), however we have also included case (a), because for this
case the coefficients $A_n$ of the associated Fourier series
(given by the representation (3.3)) can be computed explicitly.
For case (b) the coefficients $A_n$ satisfy a {\it second} order
difference equation, see equation (3.4b). We will show that this
equation has a {\it unique} solution by imposing the condition
that $A_n \rightarrow 0$ as $n \rightarrow \infty$ (which is
needed for the convergence of the series). Although we cannot give
the explicit form of $A_n$ in general, we will compute $A_n$ in
the case that $\beta$ is small. In this case we will show that the
associated series converges and that the solution $u(x,t)$ tends
to the ``inner'' and ``outer'' solutions obtained by the
perturbation expansion, see equations (5.15) and (5.16).

\section{From Burgers to the Heat Equation}

{\bf Proposition 2.1.} Let $\beta$ and $c$ be positive constants.
Let the real-valued function $u(x,t)$ solve

$\displaystyle{ \ \ \qquad \ \ \qquad  \ \ \ \ \qquad \ \ \qquad \
\ u_t = \frac{1}{2}u_{xx} - \beta uu_x + cu_x, \quad 0<x<\infty,
\quad t>0}$, \hfill (2.1a)

\vskip .1in $\displaystyle{ \ \ \qquad \ \ \qquad  \ \ \ \ \qquad
\ \ \qquad \ \ u\rightarrow 0}$ as $x \rightarrow \infty$, \hfill
(2.1b)

\vskip .1in $\displaystyle{\ \ \qquad \ \ \qquad  \ \ \ \ \qquad \
\ \qquad \ \ u(x,t)}$ is $2\pi$ periodic in $t$, \hfill (2.1c)

\vskip .1in
\noindent and one of the following boundary conditions:
$$ \int^\infty_0 u(x,t)dx = \cos t, \quad t>0, \eqno (2.2a)$$
or
$$ u(0,t) = \cos t, \quad t>0. \eqno (2.2b)$$
Define $\varphi(x,t)$ by
$$ \varphi(x,t) = e^{-\beta \int^x_\infty u(\xi,t)d\xi} - 1. \eqno (2.3)$$
Then $\varphi(x,t)$ is a real-valued function and solves

\vskip .1in $\displaystyle{ \ \ \qquad \ \ \qquad  \ \ \ \ \qquad
\ \ \qquad \ \ \varphi_t = \frac{1}{2} \varphi_{xx} + c\varphi_x,
\quad 0<x<\infty, \quad t>0}$, \hfill (2.4a)

\vskip .1in $\displaystyle{ \ \ \qquad \ \ \qquad  \ \ \ \ \qquad
\ \ \qquad \ \ \varphi \rightarrow 0}$ as $x\rightarrow \infty$,
\hfill (2.4b)

\vskip .1in $\displaystyle{ \ \ \qquad \ \ \qquad  \ \ \ \ \qquad
\ \ \qquad \ \ \varphi(x,t)}$ is $2\pi$ periodic in $t$, \hfill
(2.4c)

\vskip .1in \noindent and satisfies one of the following boundary
conditions, respectively:
$$ \varphi(0,t) = e^{\beta \cos t}-1, \quad t>0, \eqno (2.5a)$$
or
$$\varphi_x(0,t) + \beta \cos t\varphi(0,t) + \beta \cos t =0, \quad t>0. \eqno (2.5b)$$

\paragraph{Proof}
The function $\varphi$ is well defined and $\varphi \rightarrow 0$
as $x \rightarrow \infty$.

If $E$ is defined by
$$E \Doteq \exp \left[ - \beta \int^x_{\infty}
u(\xi,t)d\xi \right],$$
then

\vskip .1in $\displaystyle{ \ \ \qquad \ \ \qquad  \ \ \ \ \qquad
\ \ \qquad \ \ \varphi_x =-\beta uE},$

\vskip .1in $\displaystyle{ \ \ \qquad \ \ \qquad  \ \ \ \ \qquad
\ \ \qquad \ \ \varphi_{xx} =- \beta u_xE + \beta^2u^2E},$

\vskip .1in $\displaystyle{ \ \ \qquad \ \ \qquad  \ \ \ \ \qquad
\ \ \qquad \ \ \varphi_t = -\beta \int^x_\infty u_td\xi},$ \\
\noindent thus

\vskip .1in $\displaystyle{ \ \ \qquad \ \ \qquad  \ \ \ \ \qquad
\ \ \qquad \ \ \varphi_t - \frac{1}{2} \varphi_{xx} - c\varphi_x =
\left( \int^x_\infty u_td\xi - \frac{1}{2} u_x + \frac{1}{2} \beta
u^2-cu\right) E \beta =0}.$ \\
\noindent Also
$$\varphi
(x,t+2\pi) = e^{-\beta \int^x_\infty u(\xi,t+2\pi)d\xi} -1 =
\varphi(x,t).$$
Equation (2.2a) implies
$$ \varphi(0,t) = e^{\beta \cos t}-1.$$
Furthermore,
$$\varphi_x(0,t) =- \beta u(0,t) \exp\left[ -\beta
\int^0_\infty u(\xi,t)d\xi\right],$$
$$\varphi (0,t) = \exp \left[ -\beta\int^0_\infty
u(\xi,t)d\xi\right],$$ hence
$$ \varphi_x(0,t) = -\beta u(0,t)\varphi(0,t),$$
which implies (2.5b).

\section{The Solution for $u(x,t)$}

{\bf Proposition 3.1.} Assume that the positive constants $\beta$
and $c$, are such that $\lambda_n$ defined by
$$ \lambda_n \Doteq -c + (c^2+2in)^{\frac{1}{2}}, \quad n = 1,2,\cdots \eqno (3.1)$$
has 1 value with Re$\lambda_n <0$.  Then the solution of
(2.1)-(2.2) is given by
$$ u(x,t) = -\frac{1}{\beta} \frac{\varphi_x(x,t)}{1+\varphi(x,t)}, \eqno (3.2)$$
where
$$ \varphi(x,t) = \frac{1}{2\pi} \left\{ A_0e^{\lambda_0x} + \sum^\infty_1 \left(A_ne^{\lambda_nx +int} + {\bar A_n} e^{-{\bar
\lambda_n}
x -int} \right)\right\}, \eqno (3.3)$$ and $\lambda_0, \{
A_j\}^\infty_0$ are defined as follows:
$$ \lambda_0 =-2c;$$
in the case of the boundary condition (2.2a)
$$ A_n = \int^{2\pi}_0 e^{-int} \left[ e^{\beta \cos t}-1\right] dt, \eqno (3.4a)$$
whereas in the case of the boundary condition (2.2b)
$$ 2\lambda_n A_n + \beta(A_{n+1} + A_{n-1}) + 2\pi\beta\delta_{n,1} =0, \quad n =0,1,2,\cdots. \eqno (3.4b)$$

\paragraph{Proof.}

Equation (2.3) implies equation (3.2).  Thus the problem reduces
to solving an initial-boundary value problem for the heat equation
(2.4)-(2.5).

Let
$$ \hat\varphi(x,n) =\int^{2\pi}_0\varphi(x,t)e^{-int}dt, \eqno (3.5)$$
$$ \varphi(x,t) = \frac{1}{2\pi} \sum^\infty_{n=-\infty} \hat\varphi(x,n) e^{int} = \frac{1}{2\pi} \left\{ \hat \varphi(x,0) + \sum^\infty_1 \left(\hat\varphi (x,n)e^{int} + \hat\varphi(x,-n)e^{-int}\right)\right\}.$$
Equation (3.5) and reality imply
$$ \hat\varphi(x,-n) = \overline{\hat \varphi(x,n)}$$.

Thus
$$ \varphi(x,t) = \frac{1}{2\pi} \left\{ \hat\varphi(x,0) + \sum^\infty_1 (\hat\varphi(x,n)e^{int} + \overline{\hat \varphi(x,n)}e^{-int})\right\}. \eqno (3.6)$$
Equations (3.5) and (2.4a) imply
$$ \hat\varphi_{xx} + 2c\hat\varphi_x = 2\int^{2\pi}_0 \varphi_t(x,t)e^{-int}dt = 2\left[\left. \varphi(x,t)e^{-int}\right|^{2\pi}_0 + in\hat\varphi\right].$$
Thus using periodicity, we find
$$\hat \varphi_{xx} + 2c\hat\varphi_x - 2in\hat\varphi = 0. \eqno (3.7)$$
\paragraph{$n=0$:}
$$\hat\varphi = c_1 + A_0e^{\lambda_0x}, \quad \lambda_0 =-2c;$$
the boundness requirement as $x\rightarrow \infty$ implies
$c_1=0$. Thus
$$ \hat\varphi(x,0) = A_0e^{\lambda_0x}. \eqno (3.8a)$$
\paragraph{$n=1,2,\cdots$:}
$$ \hat\varphi(x,n) = A_ne^{\lambda_nx}, \quad \lambda^2_n + 2c\lambda_n - 2in =0. \eqno (3.8b)$$
Substituting (3.8) into (3.6) we find (3.3).  If the boundary
condition (2.2a) is valid, then
$$ A_n = \hat\varphi(0,n) = \int^{2\pi}_0\varphi(0,t)e^{-int}dt, \quad n=0,1,2,\cdots,$$
which is equation (3.4a).

If the boundary condition (2.2b) is valid, then multiplyng this
equation by $e^{-int}$ and integrating from 0 to $2\pi$ we find
$$ \int^{2\pi}_0 e^{-int} \left[ 2\varphi_x(0,t) + \beta(e^{it} + e^{-it})\varphi(0,t) + \beta(e^{it} + e^{-it})\right] =0, \quad n=0,1,\cdots.$$
Thus
$$ 2\hat\varphi_x(0,n) + \beta[\hat \varphi(0,n+1)+\hat\varphi(0,n-1)] + 2\pi\beta\delta_{n,1} =0, n=0,1,\cdots.$$
Using
$$\hat\varphi_x(0,n) = \lambda_nA_n,$$
the above equation becomes (3.4b).

\section{The Boundary Value Problem (2.1)-(2.2a) with $c=\beta^2k$ as $\beta\rightarrow 0$}
Let
$$ c=\beta^2k. \eqno (4.1)$$
The solution of this problem is given by (3.2) and (3.3) where
$A_n$ is defined by (3.4a).  The integral (3.4a) can be computed
explicitly.  For simplicity we only compute it as $\beta
\rightarrow 0$:
$$ A_n = \int^{2\pi}_0 e^{-int} \left[ \beta\cos t + \frac{\beta^2}{2!} (\cos t)^2 + O(\beta^3)\right]dt, \quad n=0,1,\cdots$$
$$ = \oint_{|z|=1} \frac{1}{iz^{n+1}} \left[ \frac{\beta}{2} (z + \frac{1}{z}) + \frac{\beta^2}{8} \left( z^2 + \frac{1}{z^2} + 2\right) + O(\beta^3)\right] dz$$
$$ = \frac{\beta}{2i} \oint_{|z|=1} \frac{1}{z^{n+1}} \left( z + \frac{1}{z}\right) dz + \frac{\beta^2}{8i} \oint_{|z|=1} \frac{1}{z^{n+1}} \left( z^2 + 2 + \frac{1}{z^2}\right) dz + O(\beta^3).$$
Thus
$$ A_n = 2i\pi \left[ \frac{\beta}{2i} \delta_{n,1} + \frac{\beta^2}{8i} \left( 2\delta_{n,0} + \delta_{n,2} \right)\right] + O(\beta^3).$$
Hence
$$ A_1 = \pi\beta + O(\beta^3), \quad A_0 = \frac{\beta^2\pi}{2} + O(\beta^3), \quad A_2 = \frac{\pi\beta^2}{4} + O(\beta^3).$$
Also
$$ \lambda_n = -\sqrt{2} e^{\frac{i\pi}{4}} \sqrt{n} - k\beta^2 + O(\frac{\beta^4}{\sqrt{n}}).$$
Therefore equation (3.3) implies
$$ \varphi(x,t) = \frac{1}{2\pi} \left\{ \frac{\beta^2\pi}{2} e^{-2\beta^2kx} + \pi\beta e^{-k\beta^2x} \left[ e^{-(1+i)x+it} + e^{-(1-i)x-it}\right] \right.$$
$$ \left. + \frac{\pi\beta^2}{4}e^{-k\beta^2x} \left[ e^{-\sqrt{2}(1+i)x+2it} + e^{-\sqrt{2}(1-i)x-2it}\right] \right\} + O(\beta^3)$$

\paragraph{1. $x = O(1)$}
$$u = -\frac{[-(1+i)e^{-(1+i)x+it} - (1-i)e^{-(1-i)x-it}] + \frac{\beta}{4} [ -\sqrt{2} (1+i)e^{-\sqrt{2}(1+i)x+2it)} - \sqrt{2}(1-i) e^{-\sqrt{2}(1-i)x-2it}] + O(\beta^2)}{ 2 + \beta [e^{-(1+i)x+it} + e^{-(1-i)x-it}] + O(\beta^2)},$$
or
$$u = \frac{1}{2} \left[ (1+i) e^{-(1+i)x+it} - (1-i) e^{-(1-i)x-it} \right] + \frac{\beta}{4}
\left\{ \frac{(1+i)e^{-\sqrt{2}(1+i)x+2it}}{\sqrt{2}} + \frac{
(1-i)e^{-\sqrt{2}(1-i)x-2it}}{\sqrt{2}} \right. \eqno (4.2)$$
$$ -2e^{-2x} - (1+i)\left. e^{-2(1+i)x+2it} - (1-i)e^{-2(1-i)x-2it} \right\} + O(\beta^2)$$

\paragraph{2. $x=O(\beta^{-2})$}
$$ u = \frac{2\beta^3k}{\beta^2 + 4e^{2\beta^2kx}}. \eqno (4.3)$$

\section{The Analysis of (2.1)-(2.2b) with $c=\beta^2k$, $k\neq 1/8$, as $\beta \rightarrow 0$}

The solution of the linear homogeneous equation (3.4b), with the
requirement that
$$ A_n \rightarrow 0 \ \ {\mathrm{as}} \ \ n \rightarrow \infty \eqno (5.1)$$
is unique.  This solution is given by
$$ A_n = -2\pi \beta F_1+(-1)^nA_0F_1\cdots F_n, \eqno (5.2)$$
where $F_n$ is defined by
$$ F_n = \frac{1}{ \frac{2\lambda_n}{\beta} - \frac{1}{ \frac{2\lambda_{n+1}}{\beta} - \frac{1}{ \frac{2\lambda_{n+2}}{\beta} \cdots}}} \eqno (5.3)$$
and the real constant $A_0$ is determined from
$$ -4cA_0 + \beta(A_1 + \bar A_1) =0 . \eqno (5.4)$$
Indeed, the definition $F_n$ implies
$$ F_n = \frac{1}{\frac{2\lambda_n}{\beta} - F_{n+1}}.$$
Solving this equation for $2\lambda_n/\beta$ we find
$$ \frac{2\lambda_n}{\beta} = F_{n+1} + \frac{1}{F_n}.$$
Substituting this expression into equation (3.4b) we obtain
$$ A_{n+1} + F_{n+1}A_n + \frac{1}{F_n} (A_n + F_nA_{n-1}) = -2\pi\beta\delta_{n,1}, \quad n=0,1,2,\cdots \eqno (5.5)$$
Letting
$$G_n \Doteq A_n + F_nA_{n-1},\eqno (5.6)$$
equation (5.5) becomes
$$ G_{n+1} + \frac{1}{F_n} G_n =-2\pi\beta\delta_n,1. \eqno (5.7)$$
Since $F_n\sim \frac{\beta}{2\lambda_n}$, $n\rightarrow \infty$,
it follows that there does {\bf not} exist a homogeneous solution
of equation (5.7) that decays as $n\rightarrow \infty$.  Thus $G_n
=-2\pi\beta F_1\delta_n,1$, and equation (5.6) yields
$$A_n + F_nA_{n-1} =-2\pi\beta F_1\delta_{n,1}.$$
The unique solution of this equation is given by (5.2). Evaluating
equation (3.4b) at $n=0$ and using $A_{-1} = \bar A_1$ we find
(5.4).

Equation (5.2) implies that
$$ A_n \sim (-1)^nA_0 \left( \frac{\beta}{2}\right)^n \prod^n_1 \frac{1}{\lambda_j}, \quad n \rightarrow \infty, \eqno (5.8)$$
thus the series (3.3) converges.

\paragraph{The asymptotic behavior as $\beta \rightarrow 0$} \ \

The definition of $\lambda_n$ (equation (3.1)) together with the
requirement that Re $\lambda_n<0$, imply that if $c=\beta^2k$,
then
$$ \lambda_n =-\sqrt{n}(1+i) + O(\beta^2). \eqno (5.9)$$
Equations (5.2), (5.3) yield
$$ A_0 = a_0 + O(\beta), A_1 = \beta a_1 + O(\beta^2), A_2 = \beta^2 a_2 + O(\beta^3), \cdots . \eqno (5.10)$$
Equation (3.4b) with $n=0,1,$ becomes
$$ n=0: \quad A_1 + \bar A_1 - 4k\beta A_0 = 0 \eqno (5.11)$$
$$ n = 1: \quad 2\lambda_1 A_1 + \beta(A_2 + A_0) + 2\pi\beta =0. \eqno (5.12)$$
Substituting equations (5.10) into (5.11), (5.12), we find
$$ a_1 + \bar a_1 = 4ka_0 \eqno (5.13a)$$
$$ -2(1+i)a_1 + a_0 + 2\pi =0. \eqno (5.13b)$$
Equation (5.13b) yields
$$ a_1 = \frac{(a_0+2\pi)(1-i)}{4}. \eqno (5.14a)$$
Substituting this expression in equation (5.13a) and using that
$a_0$ is real, we find
$$ a_0 = \frac{2\pi}{8k-1}. \eqno (5.14b)$$
Substituting (5.10) into the expression for $\varphi(x,t)$
(equation (3.3)) we obtain
$$ \varphi(x,t) = \frac{1}{2\pi} \left\{ \frac{2\pi}{8k-1} e^{-2\beta^2kx} + \frac{4k\pi}{8k-1} e^{-k\beta^2x} \left[ (1-i)e^{-(1+i)x+it} + (1+i)e^{-(1-i)x-it}\right] \beta \right\} + O(\beta^2).$$

\paragraph{1. $x =O(\beta^{-2})$}
$$ u = \frac{2\beta k}{1+(8k-1)e^{2\beta^2kx}}. \eqno (5.15)$$
\paragraph{2. $x=O(1)$}
$$ u(x,t) = \frac{1}{2} \left[ e^{-(1+i)x+it} + e^{-(1-i)x-it} \right] + O(\beta). \eqno (5.16)$$

\section*{Appendix}

\paragraph{The Analysis of equation (3.4b)} \ \

Let $A_n$ satisfy the linear homogeneous difference equation
$$ A_{n+1} + \frac{2\Lambda_n}{\beta} A_n + A_{n-1} =0, \quad n=2,3,\cdots \eqno (A.1)$$
where $\beta$ is a constant and $\Lambda_n\rightarrow\infty$
monotonically as $n\rightarrow \infty$, for example $\Lambda_n =
\lambda_n$, where $\lambda_n$ is defined by equation (3.1).  We
shall show that
$$ A_n \sim \left( -\frac{\beta}{2}\right)^{n-1} \left[ c_1 \prod^{n-1}_1 \frac{\Lambda_j}{\Lambda^2_{j-1}} + c_2 \prod^{n-1}_1 \Lambda_j\right], \quad n \rightarrow \infty. \eqno (A.2)$$
Indeed, we first make the change of variables
$$ A_n= B_n \prod^{n-1}_{j=1} \Lambda^2_j. \eqno (A.3)$$
Thus equation (A.1) becomes
$$ B_{n+1} + \frac{2}{\beta \Lambda_n} B_n + \frac{B_{n-1}}{\Lambda^2_{n-1}\Lambda^2_n} = 0, \quad n = 2,3,\cdots \eqno (A.4)$$
We write this equation in matrix form,
$$ \Psi_n = V_n \Psi_{n-1}, V_n = \left( \begin{array}{ll}
0 & 1 \\ \\
- \frac{1}{\Lambda^2_{n-1}\Lambda_n^2} & - \frac{2}{\beta
\Lambda_n}
\end{array} \right), \quad \Psi_n = \left( \begin{array}{ll} B_n
\\ \\ B_{n+1} \end{array} \right). \eqno (A.5)$$
The eigenvalues
and the eigenvectors of the matrix $V_n$ are
$$ \mu^\pm_n = - \frac{1}{\beta \Lambda_n} \pm \sqrt{ \frac{1}{\beta^2\Lambda_n^2} - \frac{1}{\Lambda^2_n \Lambda^2_{n-1} }}, \quad \left(
\begin{array}{ll}
1 \\ \\ \mu^\pm_n \end{array} \right). \eqno (A.6)$$ Using the
gauge transformation
$$ \Psi_n = T_n\Phi_n, \quad T_n =
\left( \begin{array}{ll}
1 & 1 \\ \\
\mu^+_n & \mu^-_n  \end{array} \right), \eqno (A.7)$$ we find
$$ \Phi_n = M_n T_n^{-1} T_{n-1} \Phi_{n-1}, \quad M_n =
\left( \begin{array}{ll}
\mu^+_n & 0\\ \\
0 & \mu^-_n \end{array} \right). \eqno (A.8)$$ Equation (A.6a)
implies
$$ \mu^+_n = - \frac{\beta}{2\Lambda_n\Lambda_{n-1}^2} \left( 1 + O(\frac{1}{\Lambda_n^2})\right), $$
$$ \mu^-_n = - \frac{2}{\beta \Lambda_n} \left( 1 + O(\frac{1}{\Lambda_n^2})\right). \eqno (A.9)$$
These estimates and the explicit form of $T_n$ imply that
$$ T^{-1}_nT_{n-1} = I + O(\frac{1}{\Lambda_n^3}), $$
thus the WKB approximation of equation (A.8a) yields the two
equations
$$f^+_n = \mu^+_n f_{n-1}, \quad f^-_n = \mu^-_nf_{n-1}.$$
We recall that the general solution of the first order difference
equation $F_{n+1} = g_nF_n$, is given by
$$ F_n = g_1 \ldots g_{n-1}.$$
Thus to the leading order as $n\rightarrow\infty$,
$$ f^+_n = \prod^{n-1}_1 \left( - \frac{\beta}{2\Lambda_j\Lambda_{j-1}^2} \right) = (- \frac{\beta}{2})^{n-1} \prod^{n-1}_1 \frac{1}{\Lambda_j\Lambda_{j-1}^2},$$
$$ f^-_n = \prod^{n-1}_1 \left( - \frac{\beta}{2\Lambda_j}\right) = (-\frac{\beta}{2})^{n-1} \prod^{n-1}_1 \frac{1}{\Lambda_j}.$$
Hence using (A.7) and (A.3) we find (A.2).

\section*{Acknowledgments}

\noindent ASF is grateful to A.A. Kapaev for useful suggestions.

\end{document}